\documentclass[12pt]{article}

\usepackage[hypertexnames=false]{hyperref}
 \usepackage{epsfig}
 \usepackage{graphicx} 

\def\sub#1{~\newline\noindent{  \underline{#1}} }

\usepackage{color}

\begin{document}
 \title{Eddington-Born-Infeld action for dark energy and dark matter}
 \author{M\'{a}ximo Ba\~{n}ados  \\  Departamento de F\'{\i}sica,\\
 P. Universidad Cat\'{o}lica de Chile, Casilla 306, Santiago 22,Chile. \\
 {\tt maxbanados@fis.puc.cl}}

\maketitle

\begin{abstract}
We argue that Einstein gravity coupled to a Born-Infeld theory provides an attractive candidate to represent dark matter and dark energy. For cosmological models, the Born-Infeld field has an equation of state which interpolates between matter, $\omega=0$ (small times), and a cosmological constant $\omega=-1$ (large times). On galactic scales, the Born-Infeld theory predicts asymptotically flat rotation curves.
\end{abstract}

\section{Introduction and summary of results}

The existence of dark matter and dark energy is now firmly established phenomenologically \cite{Bradac,Spergel} but the theoretical understanding is far from complete.  Einstein equations require ``exotic" components in the right hand side corresponding  to about $\%96$ of the total energy density today. Understanding the microscopic nature of these extra components is one of the most challenging and important problems faced by theoretical physics at present.

Supersymmetric and exotic particles in the standard model are the best candidates for dark matter (for a review see \cite{Jungman}).  Several experiments are now being devised for a direct detection of these particles. Alternative descriptions based on modifications to gravity have also been explored with interesting results. See \cite{Sanders,Bekenstein,Moffat,Ferreira,Ferreira2,Carroll}, and references quoted therein, for some of these efforts and its consequences. For a recent review of the Einstein aether theory see \cite{Jacobson}.

The problem of dark energy is somehow more recent, although the issue of the cosmological constant has been around for a long time. The discovery of an accelerating Universe \cite{Expansion} resulted in deep changes in cosmology. The simplest explanation for this phenomena is a small positive cosmological constant, but many other possibilities have been explored (see \cite{Carroll2,Lima} for recent reviews).

~

In this paper we consider an action for general relativity coupled to a Born-Infeld theory. The Born-Infeld theory has as fundamental variable a symmetric connection  $C^\rho_{\ \mu\nu}(x)$. $C^{\mu}_{\ \nu\rho}$ has the same symmetries and transformation properties of the Christoffel symbol but is independent from it. The action is
\begin{equation}\label{I}
  I[g_{\mu\nu},C^{\mu}_{\ \nu\rho},\Psi]  = {1 \over 16 \pi G} \int d^4x \left[ \sqrt{|g_{\mu\nu}|} R  + {2 \over \alpha l^2}\sqrt{\left|g_{\mu\nu}-l^2 K_{(\mu\nu)}\right| } \right] + \int d^4x\, {\cal L}_m(\Psi,g_{\mu\nu}),
\end{equation}
where $|A_{\mu\nu}|$, for any $A_{\mu\nu}$, denotes the absolute value of the determinant of $A_{\mu\nu}$. $K_{\mu\nu}$ is the ``Ricci" curvature associated to $C^{\mu}_{\ \nu\rho }(x)$,
\begin{equation}\label{Kmn}
K_{\mu\nu} \equiv K^{\alpha}_{\ \mu\alpha\nu } \ \ \ \   \ \  (K^{\mu}_{\ \nu\, \alpha\beta} = C ^{\mu}_{\ \nu\beta,\alpha} + C^{\mu}_{\ \sigma\alpha} C^{\sigma}_{ \ \nu\beta} - [\alpha \leftrightarrow \beta]).
\end{equation}
Besides Newton's constant, the action (\ref{I}) has two extra parameters: $l$ is a length and $\alpha$ is dimensionless.  $\Psi$ denotes all baryonic fields and ${\cal L}_{{ m}}$ the baryonic Lagrangian.

The action (\ref{I}) is similar in spirit although different in interpretation to the Born-Infeld gravity action proposed by Deser and Gibbons \cite{Deser-Gibbons},
\begin{equation}\label{IDG}
I[g_{\mu\nu}]=\int \sqrt{|g_{\mu\nu} - l^2 R_{\mu\nu} + X_{\mu\nu}(R)|}
\end{equation}
and elaborated in \cite{DG2}.  As discussed in \cite{Deser-Gibbons}, the term $X_{\mu\nu}(R)$ must be chosen such that the action is free of ghost, and free of Schwarzschild-like singularities. The action (\ref{IDG}) is an action for pure gravity, and it can be seen as a natural extension to spin two of the scalar $\sqrt{|g_{\mu\nu} + \partial_\mu \phi \partial_\nu\phi| }$ and vector $\sqrt{|g_{\mu\nu} + F_{\mu\nu}|}$ Born-Infeld (BI) theories.  For the scalar and vector BI theories the equations of motion are of second order.  For the spin two theory this is not automatic and requires the addition of $X_{\mu\nu}$.

The action (\ref{I}), on the other hand, gives rise to second order equations because $K_{\mu\nu}(C)$ depends on first derivatives of the field $C^{\mu}_{\ \nu\rho }$. This action, however, is not an action for pure gravity but gravity coupled to $C^{\mu}_{\ \nu\rho }$. The equations of motion are discussed in the appendix and in Sec. \ref{Eq/Sec} below.

It is known (e.g. \cite{Fradkin-T}) that general relativity with cosmological constant is dual to Eddington's action \cite{Eddington} $I[C]\sim\int \sqrt{|K_{\mu\nu}|}$. The action (\ref{I}) can then be interpreted as general relativity interacting with its own dual field theory.

The action (\ref{I}) can also be motivated by looking at general relativity without metric \cite{B}. This interpretation will be discussed in Sec. \ref{Sec/g=0}.

Our main goal in this paper is to argue that the field $C^{\mu}_{\ \nu\rho}$ has good properties to represent dark matter and dark energy. We shall study the equations of motion following from (\ref{I}) and prove the following properties.
\begin{enumerate}
\item
For a cosmological model, there exist solutions where the expansion factor $a(t)$ behaves as $a(t) \sim e^{Ht}$ for large $t$, and as $a(t)\sim t^{2/3}$ for small $t$.
The equation of state for the fluid interpolates between $p=0$ and $p=-\rho$.    The parameters in the solution can be adjusted such that this field contributes to $\sim 23\%$ of the total matter energy density and $\sim \%73$ of vacuum energy density, as required by observations \footnote{Couplings between dark matter and energy have appeared in \cite{Comelli}, and in \cite{Bertolami} involving a Chapligyn gas.}.

\item
For a spherically symmetric configurations, the action (\ref{I}) predicts asymptotically flat rotation curves, as required by galactic dynamics.  The parameters involved in this solution can also be adjusted to deal with realistic situations.
\end{enumerate}

We would like to stress the simplicity of this proposal. The ``Born-Infeld" term is all we need to account for both dark energy and dark matter, at least for the problems described above.  More complicated tests, like lensing, fluctuations, and others will be discussed elsewhere \cite{BFS,BRR}. See also \cite{Davi}.

\section{The Equations of Motion}

\label{Eq/Sec}

\subsection{A bi-metric theory}

The fields varied in the action (\ref{I}) are the metric $g_{\mu\nu}$ and the connection $C^{\mu}_{\ \nu\rho }$.  Both fields are independent. At the level of the equations of motion, the connection $C^{\mu}_{\ \nu\rho }$ can be written in terms of a second metric $q_{\mu\nu}$. (The full action can also be written as a bi-metric theory \cite{andy}.) This action then represent a bi-metric theory if gravity. This result follows closely the structure of Eddington's theory \cite{Eddington}. We shall postpone a detailed derivation for the appendix and include here only the result.

Let $q_{\mu\nu}(x)$ be a rank two invertible symmetric tensor satisfying the metricity condition
\begin{equation}
D_\rho q_{\mu\nu}=0
\end{equation}
{\it with respect to} $C^{\mu}_{\ \nu\rho }$.  Since $C^{\mu}_{\ \nu\rho }$ is symmetric  this implies $C^{\mu}_{\ \nu\rho} = {1 \over 2} q^{\mu\alpha} ( q_{\alpha\nu,\rho} + q_{\alpha\rho,\nu} - q_{\nu,\rho,\alpha} )$, and for every $q_{\mu\nu}$ there is a unique $C^{\mu}_{\ \nu\rho }$.

The equations of motion derived from the action (\ref{I}) can be written completely in terms of $g_{\mu\nu}$ and $q_{\mu\nu}$, and take the very simple form
\begin{eqnarray}
G_{\mu\nu} &=& - {1 \over l^2} \sqrt{{q}\over g}\, g_{\mu\alpha}\,q^{\alpha\beta}\,  g_{\beta\nu} + 8\pi G\, T^{{\scriptscriptstyle (m)}}_{\ \ \mu\nu} \label{ee} \\
K_{\mu\nu} &=& {1 \over l^2}( g_{\mu\nu} + \alpha\, q_{\mu\nu})  \label{Ke}
\end{eqnarray}
$T^{{\scriptscriptstyle (m)}}_{\ \mu\nu}$ is the energy momentum tensor associated to the baryonic Lagrangian ${\cal L}_{{\scriptscriptstyle (m)}}$. $q^{\mu\nu}$ is the inverse of $q_{\mu\nu}$. The derivation of these equations is left for the appendix.

~

Equation (\ref{ee}) is the Einstein equation.  The first term in the right hand side is the contribution from the Born-Infeld action.  Our main goal will be to prove that this fluid can account for dark matter and dark energy.

\subsection{The de-Sitter solution} \label{SecdeSitter}

The de-Sitter spacetime is an exact solution to this theory.  This can be seen as follows. (The de-Sitter spacetime is expected to be relevant after matter becomes negligible so we set here $T^{(m)}_{\ \mu\nu}=0$.)

Suppose there exists solutions of the equations of motion with $R_{\mu\nu}=\Lambda g_{\mu\nu}$.  It is direct to see that this implies that both metrics must be proportional,
\begin{equation}
q_{\mu\nu}(x) = \gamma\,  g_{\mu\nu}(x)
\end{equation}
with $\gamma$ a constant. The constant $\gamma$ can be computed as follows.
Replacing in (\ref{Ke}) we derive,
\begin{equation}
R_{\mu\nu} = {1 \over l^2}\left( \gamma \alpha + 1 \right) g_{\mu\nu}.
\end{equation}
Replacing in (\ref{ee}) (with $T^{(m)}_{\mu\nu}=0$) we derive
\begin{equation}
R_{\mu\nu} = {\gamma \over l^2} g_{\mu\nu}
\end{equation}
Consistency determines $\gamma$,
\begin{equation}
\gamma = {1 \over 1-\alpha}.
\end{equation}
Thus, the Born-Infeld field can behave as a cosmological constant with the value
\begin{equation}
\Lambda = {1 \over 1-\alpha}\, {1 \over l^2}.
\end{equation}
The value $\alpha=1$ is a critical point where cosmological solutions ceases to exist.
Curiously, we shall see that a good fit for the Friedman equation requires $\alpha$ to be close, but not equal, to one.

\section{Friedman cosmological models}

The evolution equation for the scale factor in flat cosmological models is given by the Friedman equation (neglecting radiation)
\begin{equation}\label{Fr}
{\dot a^2 \over a^2} = {\Omega_{bm} + \Omega_{dm} \over a^3} + \Omega_{\Lambda}.
\end{equation}
Current values for the (relative) densities of barionic matter $\Omega_{bm}$, dark matter $\Omega_{dm}$ and vacuum energy $\Omega_\Lambda$ are,
\begin{equation}\label{Omega}
\Omega_{bm} \simeq 0.04, \ \ \ \ \ \ \  \Omega_{dm} \simeq 0.23, \ \ \ \ \ \ \Omega_\Lambda \simeq 0.73.
\end{equation}
Among the components appearing in the right hand side of (\ref{Fr}), only the $\sim 0.04$ fraction of baryonic matter is theoretically well-understood. The other $0.23+0.73=0.96$ fraction remains a great mystery.

\subsection{Goal of this section}

The goal of this section is to demonstrate that the field $C^{\mu}_{\ \nu\rho }$ behaves like dark matter for small times, and as dark energy for larger times. In other words, its equation of state evolves from $p=0$ into $p=-\rho$.  Adjusting the parameters $\alpha$ and $l$, plus initial conditions, the Born-Infeld field can account for both the $\Omega_{dm}$ and $\Omega_{\Lambda}$ contributions in (\ref{Fr}). Thus, the action (\ref{I}), is capable to reproduce the correct evolution of the scale factor without adding neither dark matter nor dark energy.

Our approach does not shed any light into the particular values for $\Omega_\Lambda,\Omega_{dm},\Omega_{bm}$ and other cosmological parameters. We shall only prove that $l$ and $\alpha$ can be chosen such that the predictions from (\ref{I}) are consistent with the Friedman equation (\ref{Fr}).   In particular we have chosen here to set $k=0$ and consider only flat models. There is no particular reason for the choice other than simplicity. A full analysis with a varying $k$ and including other developments will be reported in \cite{BFS}.

\subsection{The ansatz and equations}
\label{Equations}

To solve (\ref{ee}) and (\ref{Ke}) we assume that both $g_{\mu\nu}$ and $q_{\mu\nu}$ are homogeneous, isotropic and with flat spatial sections.  Using the gauge freedom in the time coordinate to fix $g_{tt}=-1$, the ansatz for $g_{\mu\nu}$ and $q_{\mu\nu}$ is then,
\begin{eqnarray}\label{FRW}
g_{\mu\nu}dx^\mu dx^\nu &=& -  dt^2 + a(t)^2 (dx^2 + dy^2 + dz^2), \\
q_{\mu\nu}dx^\mu dx^\nu &=& - X(t)^2 dt^2 + Y(t)^2 (dx^2 + dy^2 + dz^2)\label{qFRW}
\end{eqnarray}
where $a(t),X(t),Y(t)$ are arbitrary functions of time to be fixed by the equations of motion and initial conditions.

As usual for flat models, and to match the choice made in (\ref{Fr}), we set
\begin{equation}
a(t) |_{ \scriptscriptstyle today }=1, \ \ \ \ \ \ \ \ \ \ \  H_0 = \dot  a(t) |_{ \scriptscriptstyle today }
\end{equation}
and use $H_0$ to define a natural dimensionless time coordinate $H_0 t$.  The time coordinate in all expressions from now on refer to this choice.

Equations (\ref{ee},\ref{Ke}) for the ansatz (\ref{FRW},\ref{qFRW}) become,
\begin{eqnarray}
{\dot a^2 \over a^2} &=& {1 \over 3l^2 H_0^2 } {Y^3\over X}{1 \over a^3} + {\rho \over \rho_c} \label{F} \label{c1} \\
\left( {Y^3 \over X} \right)^. &=& 3 X Y a \dot a  \label{c2} \\
{1 \over X^2} {\dot Y^2 \over Y^2}  &=& {1 \over 3l^2 H_0^2}\left(  - {1 \over 2 X^2} + \alpha + {3 \over 2} {a^2\over Y^2}\right), \label{c3}
\end{eqnarray}
plus second order equations related to (\ref{c1}-\ref{c3}) by Bianchi identities.
We have introduced the usual notation $\rho_c = {3H_0^2 \over 8\pi G}$. $\rho$ is the baryonic matter and we  shall assume
\begin{equation}
{\rho \over \rho_c} = {\Omega_{bm} \over a^3 }.
\end{equation}

The interpretation of equations (\ref{c1}-\ref{c3}) is straightforward. Equation (\ref{F}) is the Friedman equation determining the time evolution of the scale factor $a(t)$.  The first term in the right hand side of (\ref{c1}) is the contribution from the Born-Infeld field $C^{\mu}_{\ \nu\rho}$. Defining the density and pressure for the Born-Infeld field,
\begin{equation}\label{rhop}
\rho_{{\scriptscriptstyle BI}} = {1 \over 8\pi G l^2}{Y^3 \over X} {1 \over a^3}, \ \ \ \ \ \ \ p_{{\scriptscriptstyle BI}} = - {1 \over 8\pi G l^2} {XY \over a}
\end{equation}
the right hand side of (\ref{c1}) is simply ${1 \over \rho_{c}}(\rho_{{\scriptscriptstyle BI}} + \rho)$.  Furthermore, in terms of $\rho_{{\scriptscriptstyle BI}}$ and $p_{{\scriptscriptstyle BI}}$, equation (\ref{c2}) takes the usual conservation form
\begin{equation}
(\rho_{{\scriptscriptstyle BI}} a^3)^. = - p_{{\scriptscriptstyle BI}} (a^3)^. .
\end{equation}

Eq. (\ref{c3}) (``the Friedman equation for the metric $q_{\mu\nu}$") provides the equation of state for $\rho_{{\scriptscriptstyle BI}}$ and $p_{{\scriptscriptstyle BI}}$ allowing a full solution to the problem.  Note that using (\ref{rhop}) the functions $X(t),Y(t)$ can be written in terms of $\rho_{{\scriptscriptstyle BI}}(t),p_{{\scriptscriptstyle BI}}(t)$ and (\ref{c3}) becomes a (differential) relation between these two functions. This equation of state thus have one free parameter represented as an initial condition.

We shall now show that $\rho_{{\scriptscriptstyle BI}}$ behaves like dark matter for small times, and like dark energy for large times.

\subsection{Asymptotic $a \rightarrow 0$ and $a \rightarrow \infty$ behavior}

Due to the complicated and non-linear character of equations (\ref{c1}-\ref{c3}) we shall study them by series expansions and numerically.

~

We first  study the behavior for large values of $a$. In this regime, the baryonic matter density $\rho \sim a^{-3}$ does not contribute. (A radiation component would not contribute either.)  Neglecting the term $\rho/\rho_c$, it is direct to see that the functions,
\begin{equation}
a(t) = a_0\, e^{t/C }, \ \ \ \ \    X(t) = {1 \over \sqrt{1-\alpha}} \ \ \ \ \  Y(t) = {a_0 \over \sqrt{1-\alpha}} \, e^{t/C},
\end{equation}
with $C = \sqrt{3(1-\alpha)}\,l H_0$ provides an exact solution to (\ref{c1}-\ref{c3}). Thus, de-Sitter\footnote{The existence of this exact solution is not at all surprising because  we already know that the general equations (\ref{ee}) and (\ref{Ke}) accepts solutions of the form $R_{\mu\nu} = \Lambda g_{\mu\nu}$ when $q_{\mu\nu}$ is proportional to $g_{\mu\nu}$} space is a solution to (\ref{c1}-\ref{c3}) for large times. The constant $C$ measures the value of the associated vacuum density. In order for this solution to approach de-Sitter space with the correct exponent, we must impose
\begin{equation}\label{LO}
{1 \over 3(1-\alpha)l^2 H_0^2} = \Omega_{\Lambda}.
\end{equation}
$H_0$ and $\Omega_\Lambda$ are determined by observations. This provides a first constraint on the parameters $l$ and $\alpha$ entering in the action. We shall use (\ref{LO}) to solve $l$ in terms of $\alpha$.

~

Now, we study the $a(t) \simeq 0$ region. In this regime, an exact solution is not available, but one can display a series expansion with the desired properties. The following series
\begin{equation}
a(t)= a_0\, t^{2/3} (1 + {\cal O}(t^{4/3})), \ \ \ \ \ X(t) = x_0^3( 1 + {\cal O}(t)), \ \ \ \ \ Y(t) = x_0 (1 + {\cal O}(t) )
\end{equation}
provide a solution to (\ref{c1}-\ref{c3}). The crucial point here is the exponent $t^{2/3}$ in $a(t)$ meaning that $C^{\mu}_{\ \nu\rho}$ does indeed behave like matter for small times. The amount of dark matter is controlled by $a_0$.

\subsection{Numerical interpolation}

Our final goal is to display a solution for $a(t)$ interpolating between $a(t)\simeq t^{2/3}$ for small $a(t)$ and $a(t)\simeq e^{Ht}$ for large $a(t)$.  Furthermore, we would like this solution to exhibit the right amount of dark matter and dark energy.
This will be done by a numerical analysis.

Equations (\ref{c1}-\ref{c3}) are of first order and thus we need to give three conditions $a_1=a(1)$, $X_1=X(1)$ and $Y_1=Y(1)$, plus the values of $\alpha$ and $l$ to integrate them.  These are 5 parameters. However only two of them are independent. This can be seen as follows.

First of all, for a flat model, we can choose $a(1)=1$. Second, in (\ref{LO}), we already encounter one condition on the parameters to achieve the right evolution. Eq. (\ref{LO}) allows to solve $l$ in terms of $\alpha$. One extra condition follows by evaluating Eq. (\ref{c1}) today,
\begin{equation}
1 = {1 \over 3 l^2 H_0^2} { Y_1^3 \over X_1}  +  \Omega_{bm},
\end{equation}
from where we can solve $X_1$ in terms of $Y_1$ and $l$.  The remaining parameters are thus $\alpha$ and $Y_1$.

~

We have integrated (\ref{c1}-\ref{c3}) numerically varying $\alpha$ and $Y_1$.  The resulting curve is compared with the evolution predicted by (\ref{Fr},\ref{Omega}). Our conclusions are the following.

\begin{enumerate}

\item
First of all, there exists values of $\alpha,Y_1$ such that the evolution predicted by (\ref{Fr}) is almost undistinguishable from that following from (\ref{c1}-\ref{c3}), at least for the part of the Universe we can observe $0<t<1$. In Fig. \ref{fig1}, the continuous line represents the Born-Infeld theory with $\alpha=0.99$ and $Y_1 = 10.59$. The dots represents the Friedman evolution dictated by (\ref{Fr}).
\begin{figure}[h]
\centerline{\psfig{file=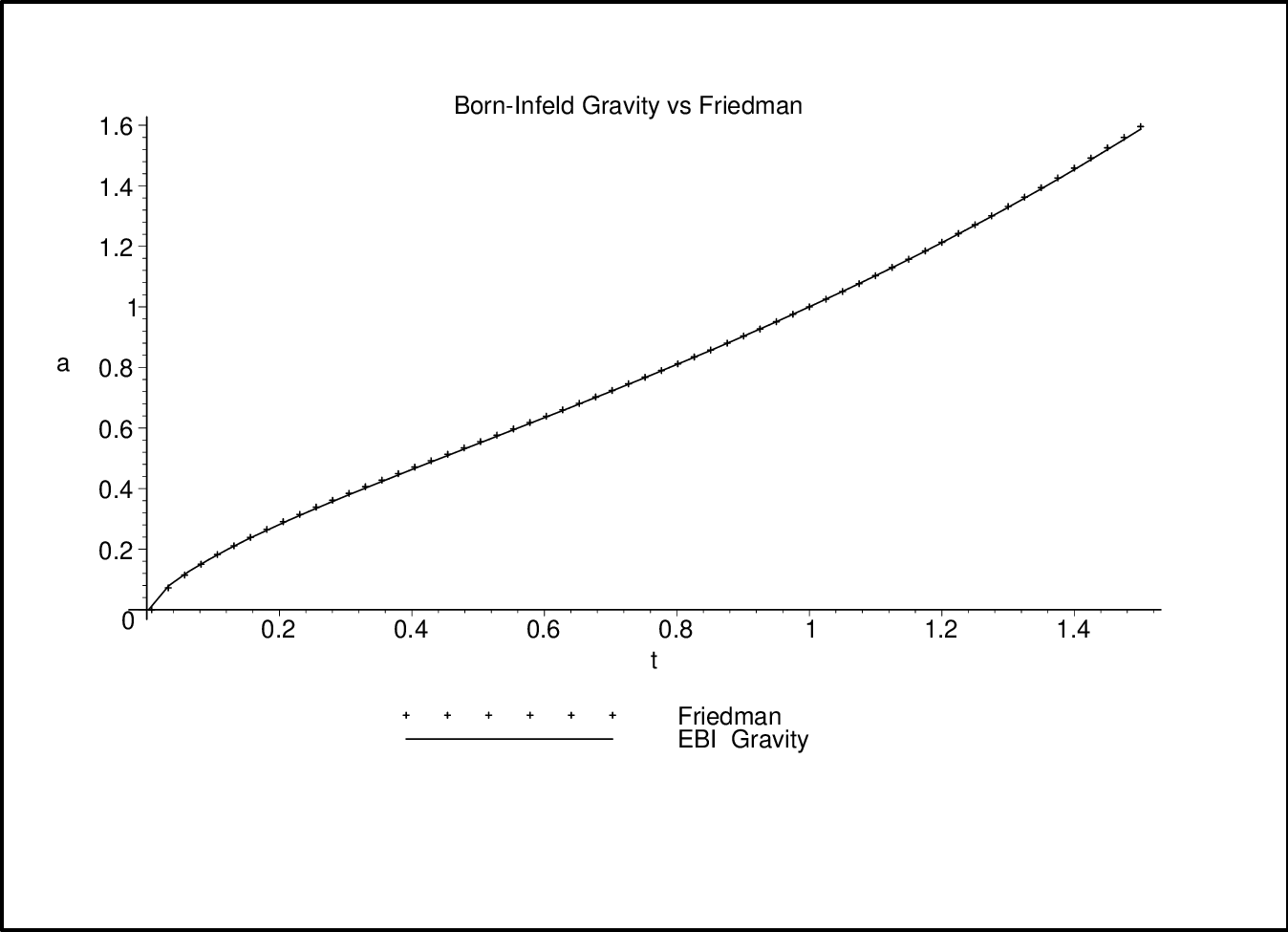,width=6cm,angle=-90}}
\caption{ \ $\alpha=0.99, \ Y_1=10.59$}
\label{fig1}
\end{figure}
The parameters were adjusted for a best fit, and in particular the big-bang occurs at $t=0.00731..$ in both theories.

\item  If $\alpha$ is not close to one, there is no value of $Y_1$ to achieve a good fit with the Friedman equation.  The above picture corresponds to $\alpha=0.99$. The value $\alpha=0.9$ also gives a good fit, but smaller ones do not. $\alpha>1$ does not work either.

The fact that $\alpha \sim 1$ to have a good fit is quite peculiar because the actual value $\alpha=1$ is singular and the de-Sitter solution does not exist (See Sec. \ref{SecdeSitter}).  In any case, recall that $\alpha$ enters in the action as a coupling constant and is not subject to variations.  More testings on the theory should narrow the actual value of this parameter.

\item
Of course no measurements exist for $t>1$, but it is interesting to explore the predictions of Born-Infeld theory to larger times. If one chooses the parameters such that the Big-Bang occurs at the same value of $t$ in both theories, then for large $t$ the expansion factor $a(t)$ grows slightly slower in the Born-Infeld theory.   Further details on this issue will be reported  elsewhere.
\end{enumerate}

\subsection{The evolution of the equation of state}

\begin{figure}[h]
 \centerline{\psfig{file=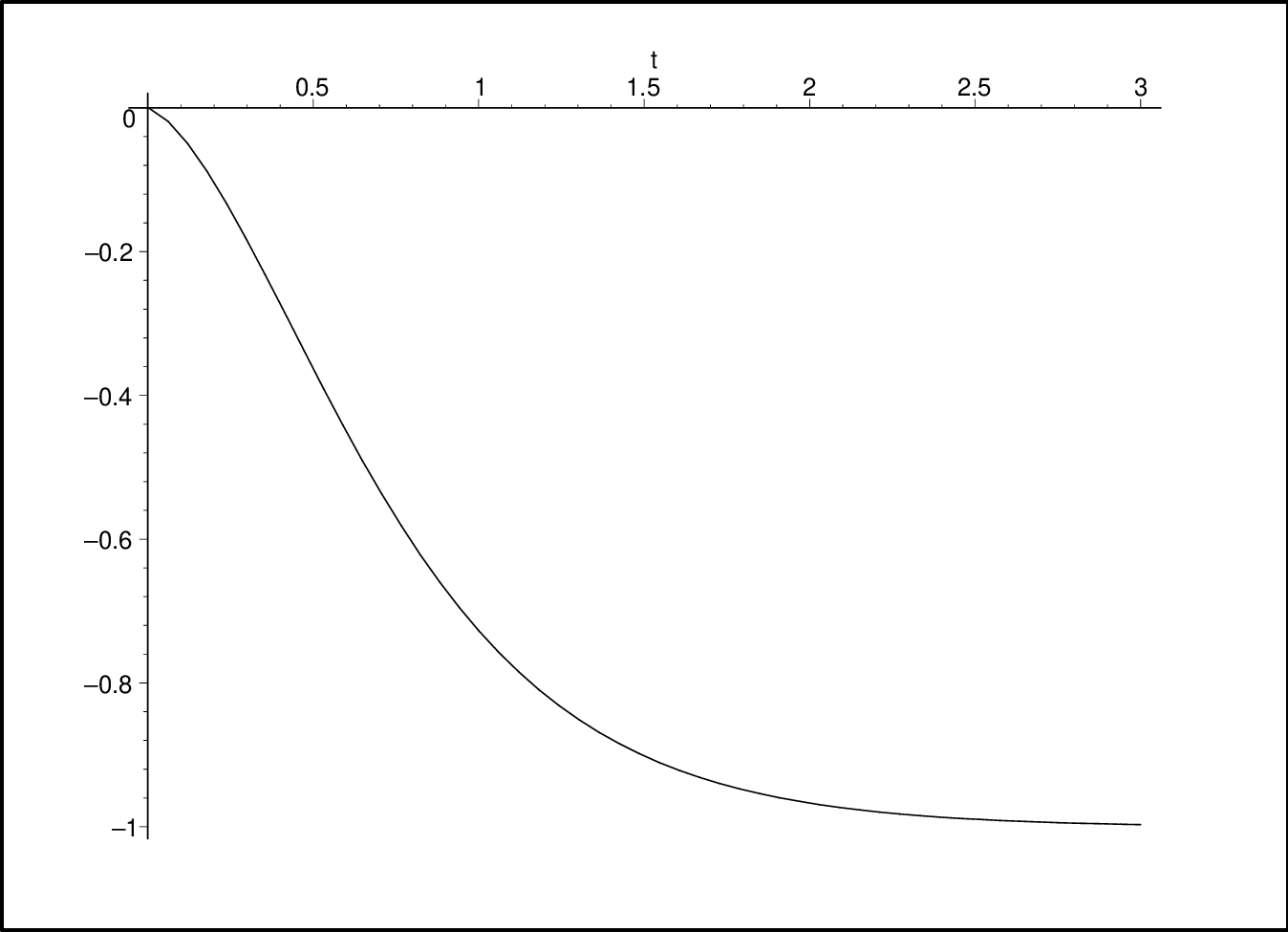,width=6cm,angle=-90 }}
\label{eqst}
\caption{ Evolution of the equation of state}
\end{figure}

As we mention in Sec. \ref{Equations}, the field $C^{\mu}_{\ \nu\rho}$ can be characterized by an energy density $\rho_{{\scriptscriptstyle BI}}$ and pressure $p_{{\scriptscriptstyle BI}}$ whose expressions are given in (\ref{rhop}).  The corresponding equation of state is,
\begin{equation}
{p_{{\scriptscriptstyle BI}} \over \rho_{{\scriptscriptstyle BI}}} = - \left( {a X \over Y} \right)^2
\end{equation}
and we observe that the pressure is always negative.  Fig. \ref{eqst} shows the evolution $0<t<3$ of the quotient $p_{{\scriptscriptstyle BI}}/\rho_{{\scriptscriptstyle BI}}$.
We see clearly the interpolation between $p_{{\scriptscriptstyle BI}}=0$ for small times, and $p_{{\scriptscriptstyle BI}}=-\rho_{{\scriptscriptstyle BI}}$ for larger times.

~

\section{Spherical symmetry and galactic rotation curves}

~

\noindent \underline{NOTE:} {\bf Equations (45)-(46) contain a sign error, which invalidates
the remaining analysis (thanks to A. Reisenegger for pointing this out). The problem arises from a mistake when calculating the absolute value $|\det g/\det q|$. The author
still believes that this theory can account for DM on galactic scales,
but requires further study. We hope to come back to this point in the near future.
The preceding analysis of cosmological scales is free of errors.}

~

~

In this section we explore the action (\ref{I}) on galactic scales where dark matter also plays an important role. (Dark energy is less relevant at this scale.) Stars orbiting galaxies have rotation curves not matching the observed luminous matter and this implies the existence of `dark matter'. See \cite{Rubin} for a recent review. If the Born-Infeld theory (\ref{I}) can be regarded as a good candidate for this exotic form of matter, it must account for these flat rotation curves.

Numerical studies of matter interacting only gravitationally suggests the 2-parameter NFW \cite{NFW} density profile for the dark matter halo,
\begin{equation}\label{NFWp}
\rho_{{\scriptscriptstyle NFW}}(r) = {a \over r( 1 + r/r_0)^2}.
\end{equation}
The density diverges at the origin but the total mass is finite. The rotation curve associated to this profile has a peak and decreases slowly with $r$.

Another popular density function is the (pseudo) Isothermic profile,
\begin{equation}\label{ISOp}
\rho_{{\scriptscriptstyle ISO}}(r) = {b \over 1 + r^2/r_0^2}
\end{equation}
having a finite density at the origin. This halo leads to a rotation curve increasing  monotonically with an asymptotically flat region. See \cite{deBlok} for an observational comparison of both profiles.

~

Our proposal is that the Born-Infeld field appearing in (\ref{I}) can account for the dark matter present in galaxies. We shall prove that the action (\ref{I}) gives rise to a rotation curve behaving like the NFW halo for small $r$, and as the (pseudo) Isothermal halo for large $r$.  In particular, the Born-Infeld theory yields an asymptotically flat rotation curve (at least to first order in the coupling constant).  To this end, we study in this section solutions with spherical symmetry to equations (\ref{ee}-\ref{Ke}).

\subsection{The ansatz and equations of motion}

Our main interest is the `dark matter' contribution to the gravitational potential induced by $C^{\mu}_{\ \nu\rho}$.   We shall then neglect here the visible matter, set $T^{{\scriptscriptstyle (m)}}_{\ \ \mu\nu}=0$ and study the solutions of
\begin{eqnarray}
G_{\mu\nu} &=& - {1 \over l^2} \sqrt{{q}\over g}\, g_{\mu\alpha}\,q^{\alpha\beta}\,  g_{\beta\nu}\, \label{oee} \\
K_{\mu\nu} &=& {1 \over l^2}( g_{\mu\nu} + \alpha\, q_{\mu\nu})  \label{oKe}
\end{eqnarray}
with spherical symmetry.

First, note that since $g_{\mu\nu}$ and $q_{\mu\nu}$ are invertible, the right hand side of (\ref{oee}) is different from zero at all points.  This represents a non-compact source.  Since for a galactic problem we expect the curvature to be very small, the right hand side of (\ref{oee}) must be very small.  This is indeed true because
the parameter $l$ is of cosmological scale, and thus ${1 \over l^2}$ is very small compared to any galactic scale.

This also means that we can treat the right hand side of (\ref{oee}) and (\ref{oKe}) as perturbations.  The parameter ${1 \over l^2}$ measures the interaction between $g_{\mu\nu}$ and $q_{\mu\nu}$. If ${1 \over l^2}\rightarrow 0$, these two fields are decoupled and satisfy the order zero equations,
\begin{equation}\label{zero}
G_{\mu\nu}=0, \ \ \ \ \ \  K_{\mu\nu}=0
\end{equation}

The interactions can be incorporated perturbatively by expansions of the form,
\begin{eqnarray}
g_{\mu\nu} &=& g_{\mu\nu}^{\scriptscriptstyle{(0)}} + {1 \over l^2 } \, g_{\mu\nu}^{\scriptscriptstyle{(1)}}  + {1 \over l^4 } \, g_{\mu\nu}^{\scriptscriptstyle{(2)}} + \cdots\\
q_{\mu\nu} &=& q_{\mu\nu}^{\scriptscriptstyle{(0)}} + {1 \over l^2 } \,q_{\mu\nu}^{\scriptscriptstyle{(1)}}  + {1 \over l^4}\,  q_{\mu\nu}^{\scriptscriptstyle{(2)}}\, + \cdots.
\end{eqnarray}
where $g^{{\scriptscriptstyle (0)}}_{\mu\nu}$ and $q^{{\scriptscriptstyle (0)}}_{\mu\nu}$ satisfy the order zero equations (\ref{zero}). [There exists a different perturbative scheme leading to a different sector of this theory. Instead of treating ${1 \over l^2}$ as very small, one could start with the exact solution described in Sec. \ref{SecdeSitter}, and study linearized fluctuations around that background. This yields a different set of solutions that will be studied elsewhere.]

~

We start our discussion with the order zero fields  $g^{{\scriptscriptstyle (0)}}_{\mu\nu}$ and $q^{{\scriptscriptstyle (0)}}_{\mu\nu}$. The static, spherically symmetric solution to the order zero equation $G_{\mu\nu}=0$ is the Schwarzschild metric
\begin{equation}\label{g0}
g_{\mu\nu}^{\scriptscriptstyle{(0)}}\, dx^\mu dx^\nu  = -c^2\left( 1 - {2MG \over c^2 r} \right)  dt^2 +  \left( 1 - {2MG \over c^2 r} \right)^{-1} dr^2 + r^2d\Omega^2.
\end{equation}
However, $M$ represents the total baryonic mass and since we have set $T^{{\scriptscriptstyle (m)}}_{\ \ \mu\nu}=0$ we also set
\begin{equation}
M=0.
\end{equation}
Our order zero $g^{{\scriptscriptstyle (0)}}_{\ \mu\nu}$ metric is thus flat space. The effects of baryonic matter can easily be incorporated at the end and will be studied in \cite{BRR}.

The order zero equation for $q_{\mu\nu}$ is $K_{\mu\nu}=0$. The solution with spherical symmetry is also the Schwarzschild metric in the ``reciprocal" metric $q_{\mu\nu}$,
\begin{equation}\label{q0}
q_{\mu\nu}^{\scriptscriptstyle{(0)}}\, dx^\mu dx^\nu    = -\beta^2 c^2\left( 1 - {w_0 \over  \tilde k(r)} \right) dt^2 + \left( 1 - {w_0 \over \tilde k(r)} \right)^{-1}  \tilde k'^2 dr^2  +  \tilde k^2(r) d\Omega^2.
\end{equation}
Here, $\tilde k(r)$ is an arbitrary function of $r$, $w_{0}$ is an arbitrary constant with dimensions of length (the $q_{\mu\nu}$ ``Schwarzschild radius"), and $\beta$ is a dimensionless constant.  This tensor solves $K_{\mu\nu}=0$ everywhere except at $\tilde k=0$. Some comments on the metric (\ref{q0}) are in order:

\begin{enumerate}
\item
We have chosen the radial coordinate $r$ such that $g^{{\scriptscriptstyle (0)}}_{\mu\nu}$ has $r^2d\Omega^2$ in the angular part. It is then {\it not} correct to assume that the metric $q^{{\scriptscriptstyle (0)}}_{\mu\nu}$ can also be written in terms of $r$ and with the same $r^2 d\Omega^2$ in its angular part. This is role of the function $\tilde k(r)$ which represents an arbitrary radial re-parametrization.  This function will be determined by the equations of motion.

\item
In the same way, the time scale $t$ is fixed in terms of the metric $g_{\mu\nu}$, and does not need to be the same for the metric $q_{\mu\nu}$. This the role of the dimensionless constant $\beta$ entering in (\ref{q0}). This constant will be important below.

\item
If $w_0\neq 0$, the metric (\ref{q0}) solves $K_{\mu\nu}=0$ everywhere, except at $k=0$. As we shall see below, the most interesting solution requires $w_0\neq 0$, and explores $k(r)$ all the way to the horizon. This means that we are forced to interpret the metric (\ref{q0}) as a black hole in the $q_{\mu\nu}$ space.  We shall restrict our discussion to the region,
\begin{equation}
\tilde k > w_0.
\end{equation}
\item
For $w_0\neq 0$ it will be convenient to use a dimensionless radial coordinate,
\begin{equation}
k \equiv {\tilde k \over w_0}.
\end{equation}
In particular the horizon is now located at,
\begin{equation}
k=1, \ \ \ \ \ (\mbox{horizon}).
\end{equation}
From now on, all formulas refer to this coordinate.

\end{enumerate}

~

Having chosen the zero order solutions to (\ref{oee}) and (\ref{oKe}), we now discuss the corrections induced but the right hand side of these equations. We only discuss here the first order correction to $g_{\mu\nu}$, proportional to ${1 \over l^2}$. Since the right hand side of (\ref{oee}) is already of order ${1 \over l^2}$, it is enough to know $q_{\mu\nu}$ to order zero. [Note that $q_{\mu\nu}^{{\scriptscriptstyle (0)}}$ contributes to $g_{\mu\nu}^{{\scriptscriptstyle (1)}}$, $q_{\mu\nu}^{{\scriptscriptstyle (1)}}$ contributes to $g_{\mu\nu}^{{\scriptscriptstyle (2)}}$, and so on.]

Our problem then reduces to replacing $q_{\mu\nu}$ given by (\ref{q0}) in (\ref{oee}) and solve for the metric $g_{\mu\nu}$ to first order in ${1 \over l^2}$. The metric $g_{\mu\nu}$ must be spherically symmetric.  We then write,
\begin{eqnarray}\label{metric}
ds^2 &=& -c^2\left( 1 + {1 \over c^2}\Phi(r)\right) dt^2 +\left(1 - {2m(r) \over c^2 r} \right)^{-1} dr^2  + r^2d\Omega^2,
\end{eqnarray}
with
\begin{eqnarray}
  \Phi &=& \Phi^{{\scriptscriptstyle (0)}} + {1 \over l^2}\, \Phi^{{\scriptscriptstyle (1)}} + {1 \over l^4}\, \Phi^{{\scriptscriptstyle (2)}} + \cdots \\
   m &=& m^{{\scriptscriptstyle (0)}} + {1 \over l^2}\, m^{{\scriptscriptstyle (1)}}  + {1 \over l^4}\, m^{{\scriptscriptstyle (2)}} + \cdots.
\end{eqnarray}
As we have already discussed, in the approximation with no baryonic matter, the zero order solution is simply flat space and thus
\begin{equation}
 \Phi^{{\scriptscriptstyle (0)}}=0, \ \ \ \  m^{{\scriptscriptstyle (0)}}=0.
\end{equation}

To first order we obtain the equations,
\begin{eqnarray}
    {dm^{{\scriptscriptstyle (1)}} \over dr} \left(1 - {1 \over k}\right) - {w_0^3  c_0^2 \over 2 \beta}\, k^2{d k \over dr} &=& 0, \label{g11}\\
  \beta c^2w_0 k^2\left(1 - {1 \over k}\right)  + 2 {dk\over dr} u^{{\scriptscriptstyle (1)}} &=& 0, \label{g22} \\
  {du^{{\scriptscriptstyle (1)}} \over dr} + \beta c^2 w_0\,r\, {dk \over dr}  &=& 0, \label{g33}
\end{eqnarray}
where we have re-defined $\Phi^{{\scriptscriptstyle (1)}}(r)$ in terms of a new function $u^{{\scriptscriptstyle (1)}}(r)$ by
\begin{equation}\label{phi1}
r{d\Phi^{{\scriptscriptstyle (1)}} \over dr}=   u^{{\scriptscriptstyle (1)}}(r) + {m^{{\scriptscriptstyle (1)}}(r) \over r}.
\end{equation}
[To first order, the equations only depend on $\Phi'$ and this is why this redefinition does not spoil locality.]

\subsection{Full parametric solution. Two branches}

Equations (\ref{g11}-\ref{g33}) are three non-linear equations for the three unknowns $m^{{\scriptscriptstyle (1)}}(r),u^{{\scriptscriptstyle (1)}}(r)$ and $k(r)$.  A much simpler set of equations can be obtained by changing the independent variable from $r$ to $k$.

We define the functions $u^{{\scriptscriptstyle (1)}}(k),m^{{\scriptscriptstyle (1)}}(k)$ and $r(k)$. Also, for any $f(r)$,
\begin{equation}
{df\!(r) \over dr}  = \left. {df(k) \over dk}\right/ {dr \over dk}.
\end{equation}
Performing these substitutions,  equations (\ref{g1}-\ref{g3}) become linear for the unknowns $m^{{\scriptscriptstyle (1)}}(k),u^{{\scriptscriptstyle (1)}}(k)$ and $r(k)$,
\begin{eqnarray}
    {dm^{{\scriptscriptstyle (1)}} \over dk} \left(1 - {1 \over k}\right) - {w_0^3  c_0^2 \over 2 \beta}\, k^2  &=& 0 \label{g1}\\
  \beta c^2 w_0 k^2\left(1 - {1 \over k}\right){dr \over dk}  + 2\, u^{{\scriptscriptstyle (1)}} &=& 0 \label{g2} \\
  {du^{{\scriptscriptstyle (1)}} \over dk} + \beta c^2 w_0\,r   &=& 0 . \label{g3}
\end{eqnarray}
Note in particular that $m^{{\scriptscriptstyle (1)}}$ has decoupled from $u^{{\scriptscriptstyle (1)}}$ and $r(k)$. The general solution can be found in closed form,
\begin{eqnarray}
  r(k) &=& A_0 \left( -\left(k - {1 \over 2}\right)\ln\left(1-{1 \over k}\right) -1\right) + B_0 \left(k - {1 \over 2} \right) \label{rk} \\
  u^{{\scriptscriptstyle (1)}}(k) &=& {1 \over 2} \beta c^2 w_0 \left[ A_0 \left( k^2 \left( 1- {1 \over k} \right) \ln\left( 1 - {1 \over k}\right) +  k - {1 \over 2} \right) - B_0 ( k^2 - k) \right] \label{uk} \\
  m^{\scriptscriptstyle (1)}(k) &=& {w_0^3c^2\over 2\beta}\left( {1 \over 3}k^3 + {1 \over 2}k^2 + k + \ln(k-1) - h_0 \right) \label{mk}
\end{eqnarray}
where $A_0,B_0$ and $h_0$ are integration constants.   This solution is real for $k>1$, that is outside the horizon in the reciprocal space $q_{\mu\nu}$.

~

To explore the properties of the different solutions we first note that the function $r(k)$ displayed in (\ref{rk}) diverges at two different values of $k$,
\begin{equation}
k = \infty, \ \ \ \ \ \ \ \ \mbox{and} \ \ \  \ \ \ \ \ k=1.
\end{equation}

Since the function $r(k)$ is a coordinate change and must be globally defined at least in the range $0<r<\infty$, the derivative $dr/dk$ must be different from zero everywhere. Note that if $A_0$ and $B_0$ have the same sign, then the function $r(k)$ has a maximum or minimum, which is not allowed.  This leaves two simple cases:

\begin{itemize}
\item {\bf Linear branch:} $A_0<0,B_0>0$. In this case, $r(k)$ diverges for large $k$, and becomes zero at some finite value $k=k_0$.  For $k<k_0$ the coordinate $r(k)$ is negative and thus this region is not physical. For large $k$ the solution rapidly approaches a linear behavior. The physical range of the coordinate $k$ in this case is
\begin{equation}
k_0 \leq k < \infty.
\end{equation}
However, it can be easily seen that $m^{{\scriptscriptstyle (1)}}/r$ diverges quadratically for large $r$. This behavior is unacceptable.  The same divergency is observed for the potential $\Phi(r)$. From now on we shall exclude this case.

\item {\bf Logarithmic branch:} $A_0>0,B_0<0$. In this case, $r(k)$ diverges at $k=1$ and becomes zero at some finite value $k=k_0$. The physical range of the coordinate $k$ in this case is
\begin{equation}
1 > k \geq k_0
\end{equation}
The most salient and peculiar property of this branch is that infinity is mapped to the horizon in the metric $q_{\mu\nu}$. There is a strong/weak relationship between both fields. The details of this branch are studied in the following paragraphs.

\end{itemize}

\begin{figure}[h]
\centerline{\psfig{file=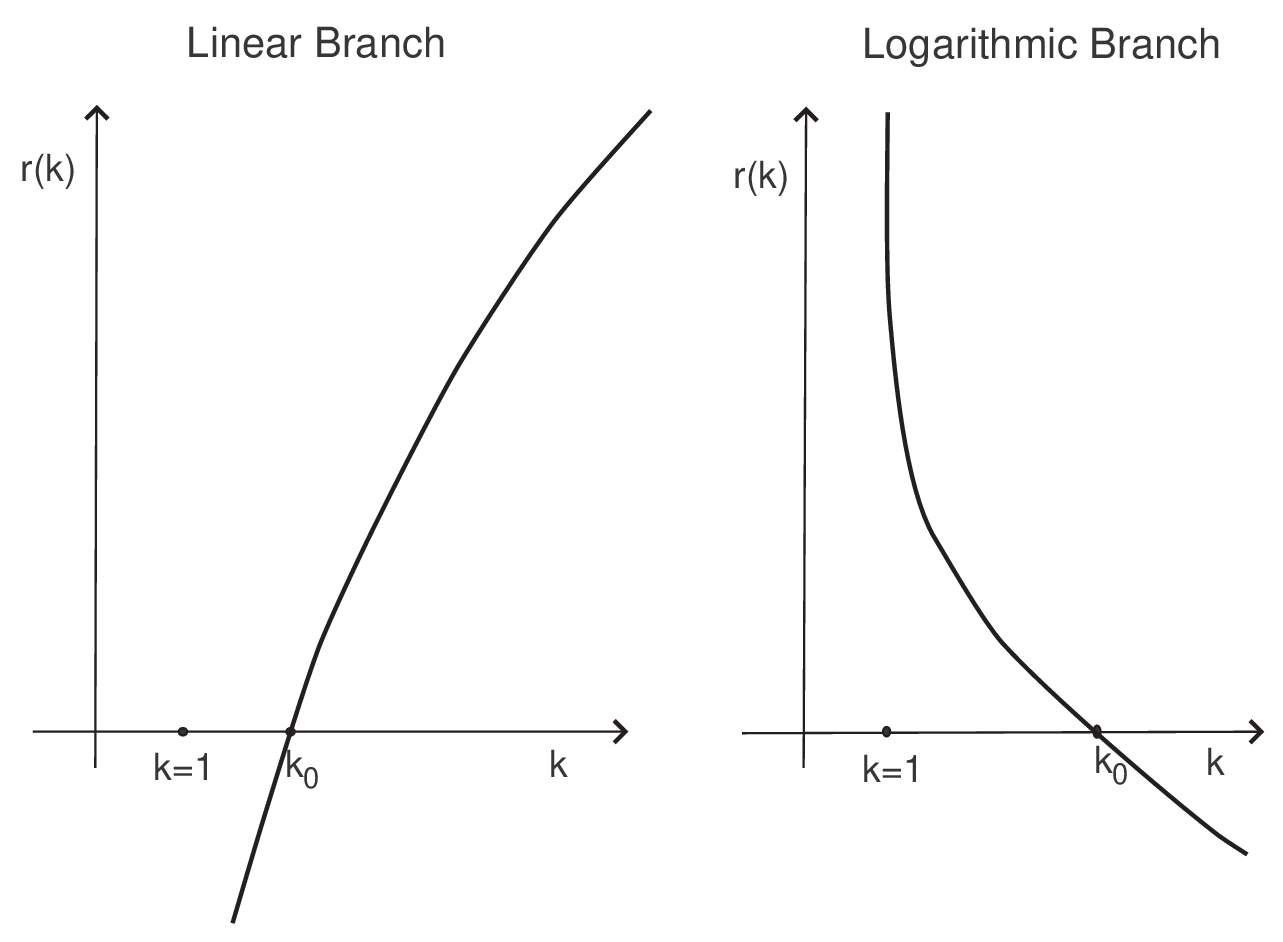,width=10cm,angle=0}}
\caption{Two branches}
\label{bran}
\end{figure}

Fig. \ref{bran} shows the behavior of the function $r(k)$ for each branch.

\subsection{The logarithmic branch and asymptotically flat rotation curves}

The most important property of this branch is that the rotation curves are asymptotically flat.  Let us recall the relation between the Newtonian potential appearing in (\ref{metric}) and the rotation speed of a (non-relativistic) object at distance $r$,
\begin{equation}\label{prof}
v(r) = \sqrt{ r {d\Phi(r) \over dr}}.
\end{equation}
(This follows from the geodesic equation.) On the other hand, the derivative of the potential $\Phi$, to first order in ${1 \over l^2}$, is given in terms of $u^{{\scriptscriptstyle (1)}}$ and $m^{{\scriptscriptstyle (1)}}$ in (\ref{phi1}).  The rotation curve can be expressed as a parametric function ,
\begin{equation}
v(k) = {1 \over l} \sqrt{ u^{{\scriptscriptstyle (1)}}(k) + {m^{{\scriptscriptstyle (1)}}(k) \over r(k)}}, \ \ \ \ \ \   r=r(k)
\end{equation}
where $u^{{\scriptscriptstyle (1)}}(k),m^{{\scriptscriptstyle (1)}}(k)$ and $r(k)$ are given in (\ref{rk}-\ref{mk}).

From these expression is it direct to compute the limit,
\begin{eqnarray}\label{vinf0}
v_{\infty}^2 & \equiv & \lim_{k\rightarrow 1} v^2(k) \nonumber\\
&=&  {w_0 (\beta^2 A_0^2 - 4 w_0^2) \over 4A_0 \beta l^2}\, c^2
\end{eqnarray}
which is indeed finite.

However, this is not the whole story. We need to impose boundary conditions at $r=0$ ($k=k_{0}$) to ensure that the solution and in particular the rotation curve (\ref{prof}) is well-behaved there too. This will imply the following constraints and redefinitions of  the parameters $A_0,B_0$ and $h_0$.

\begin{enumerate}
\item
We first express $B_0$ in terms of $k_0$, the point where $r(k_0)=0$. This gives the following expression for $B_0$,
\begin{equation}
B_0 = {A_0 \over 2k_0-1} \left((2k_0-1) \ln\left(1 - {1 \over k_0} \right) + 2\right).
\end{equation}
\item
Second, ${m^{{\scriptscriptstyle (1)}}}\over r$ must be finite at $r=0$.  This implies  that $m^{{\scriptscriptstyle (1)}}(k)$ must vanish at $k=k_0$ and this fixes $h_0$ to be \begin{equation}
 h_0 = {1 \over 3} k_0^3 + {1 \over 2}k_0^2 + k_0 + \ln(k_0-1)
\end{equation}
\item
Finally, the orbital velocity of an object at $r=0$ must be zero. This implies that $u^{{\scriptscriptstyle (1)}} + m^{{\scriptscriptstyle (1)}}/r$ evaluated at $k=k_0$ must vanish. This is achieved by choosing the constant $A_0$ to be
\begin{equation}
 A_0 = {2 w_0k_0^2 (2k_0-1) \over \beta}
\end{equation}
\end{enumerate}
In summary, boundary conditions at $r=0$ fix $B_0,A_0$ and $h_0$ in terms of a new parameter $k_0$. The full solution is then characterized by three remaining constants. The length scale $w_0$, and two dimensionless numbers $\beta$ and $k_0$.

\subsection{A better parametrization and examples}

The solution we have found is still parameterized by several numbers. The functions $r(k),v(k)$ depend on $l,c,\beta,w_0,k_0$.   The first two, $l,c$ enter in the action and cannot be varied. In fact $l$ has been already constrained by the cosmological analysis.  The other three remaining parameters can be chosen to match a desired physical situation.  Before plotting examples is it convenient to choose a different basis for these three arbitrary parameters.

First, the asymptotic velocity $v_{\infty}$ computed in (\ref{vinf0}) in terms of $k_0$ is
\begin{equation}\label{vinf}
v^2_\infty  = {4k_0^6-4k_0^5+ k_0^4 - 1 \over 2(2k_0-1)k_0^2}\, {w_0^2 \over l^2}\, c^2
\end{equation}
This parameter is of course a natural observable which can be identified easily for most galaxies. We use this equation and express $w_0$ in terms of $v_\infty$,
\begin{equation}\label{w0}
w_0 = \,\sqrt{{2(2k_0-1)k_0^2 \over 4k_0^6-4k_0^5+ k_0^4 - 1}}\, {l\, v_\infty \over c}
\end{equation}
Second, the dimensionless parameter $\beta$, which enter in (\ref{q0}), can be redefined as
\begin{equation}
\beta = {l \, \over r_0} { v_\infty \over c}.
\end{equation}
where $r_0$ is an arbitrary parameter with dimensions of length.

With these definitions, the functions $r(k),v(k)$ take the convenient form
\begin{equation}
r(k) = r_0 f_1(k,k_0), \ \ \ \ \ \   v(k) = v_\infty f_2(k,k_0).
\end{equation}
The arbitrary constant $r_0$ sets the length scale while $v_\infty$ set the velocity scale. Since both are arbitrary, they can be fixed to any desired values to fit realistic curves.   The constant $k_{0}$ controls the shape of the curve and how fast it grows. Since there are three independent parameters, there will be a degeneracy when fitting these curves with observational data (this will be discussed in \cite{BRR}).
The explicit expressions for $f_1,f_2$ are not very illuminating, and can be derived directly from the solution (\ref{rk}-\ref{mk}). Of course $f_2$ satisfies $f_{2}(1,k_0)=1$.

~

\begin{figure}[h]
\centerline{\psfig{file=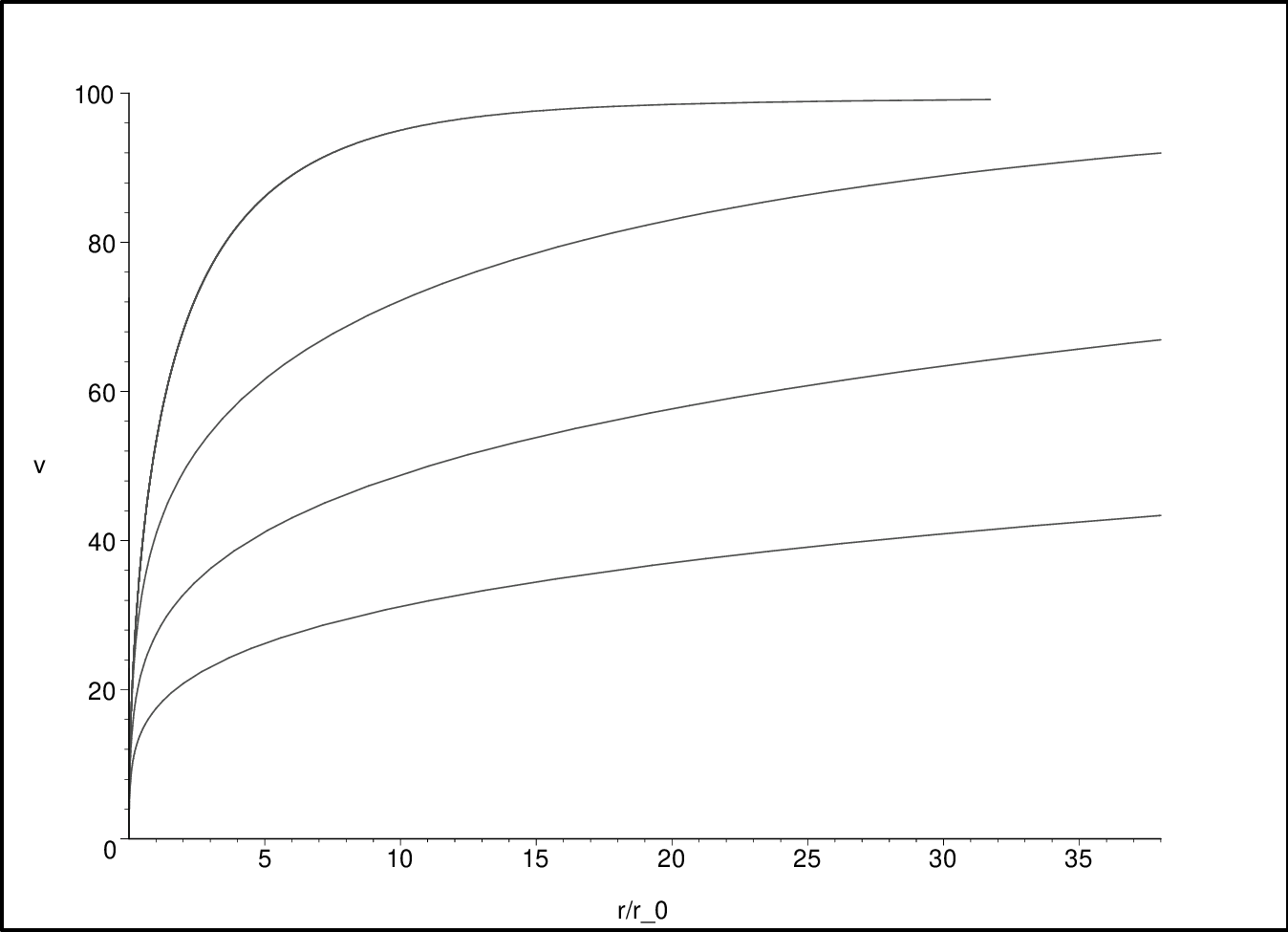,width=6cm,angle=270}}
\caption{Rotation curves for $k_0=50,15,5,1.5$.      }
\label{figg1}
\end{figure}

Fig. (\ref{figg1}) shows examples of the curve with $v_\infty=100km/sec$, $r_0$ fixed, and varying $k_{0}$. The top curve corresponds to $k_0=1.5$. As $k_0$ increases we observe a slower growth of the rotation curve. All curves asymptotically reach the value $v_\infty=100km/sec$.  The horizontal axis is expressed in terms of $r/r_0$, and choosing $r_0$ one can fit any desired length scale.

\begin{figure}[h]
\centerline{\psfig{file=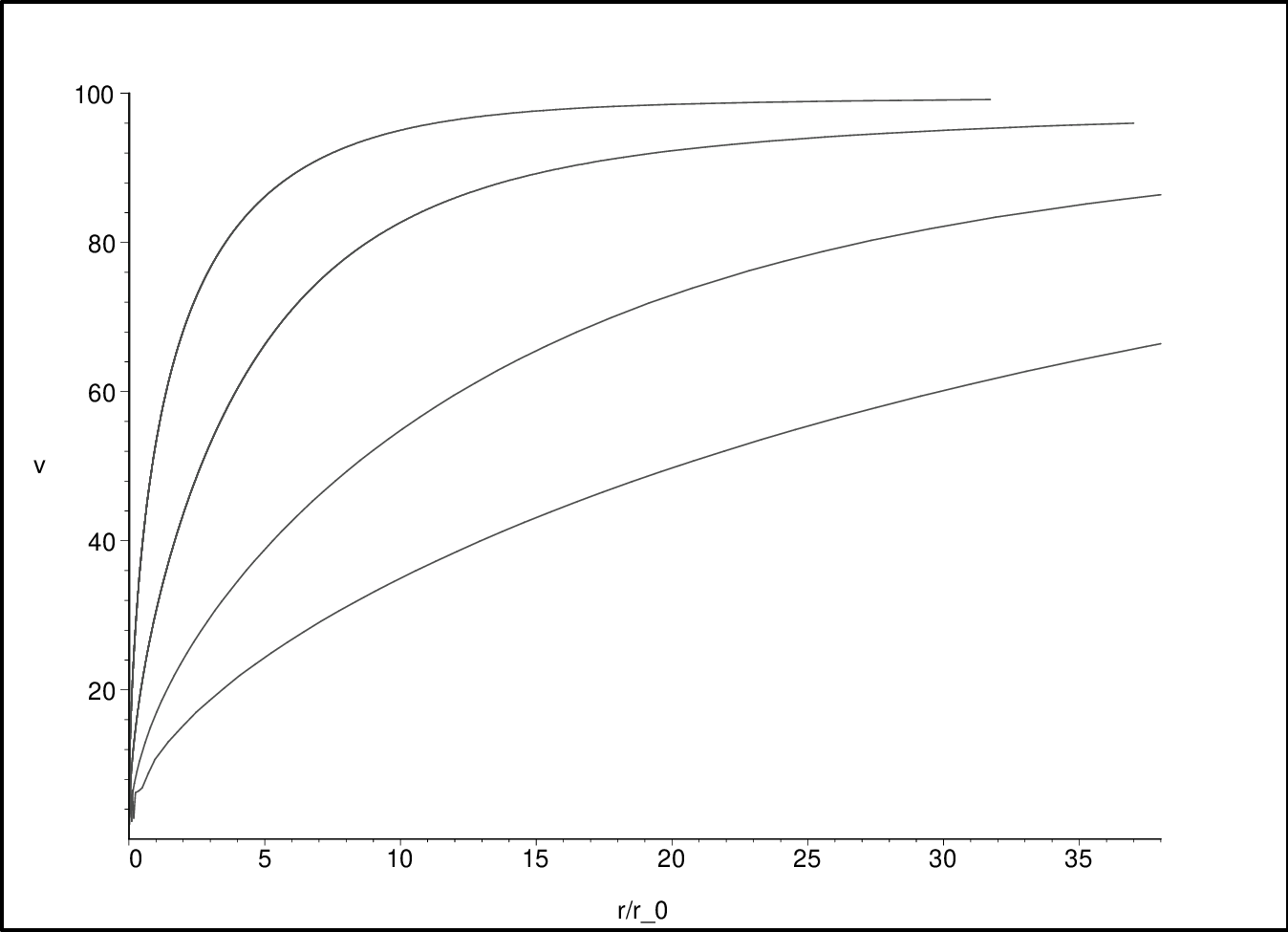,width=6cm,angle=270}}
\caption{Rotation curves for $k_0=1.5,1.03,1.003,1.0005$.      }
\label{figg2}
\end{figure}

It is interesting to note that for values of $k_0$ smaller than $k_0 \simeq 1.5$, the curves change shape.  Fig. (\ref{figg2}) shows the rotation curve for $k_0=1.5,1.03,1.003,1.0005$. The top curve corresponds to $k_{0}=1.5$. As $k_{0}$ becomes smaller, the rotation curves growths more slowly.

Note that one does not expect the curves to be asymptotically flat to all orders. The solutions discussed here are only the first order approximation in the coupling ${1 \over l^2}$. The next orders are necessary to extrapolate the result to large values of $r$, comparable with $l$.  Also, the near horizon region for the metric  $q_{\mu\nu}$ is singular in Schwarzschild  coordinates and thus a proper analysis in regular coordinates may also change the behavior near infinity.

\subsection{Final remarks}

We end this section with two extra comments regarding the solutions with spherical symmetry.

\begin{figure}[h]
\centerline{\psfig{file=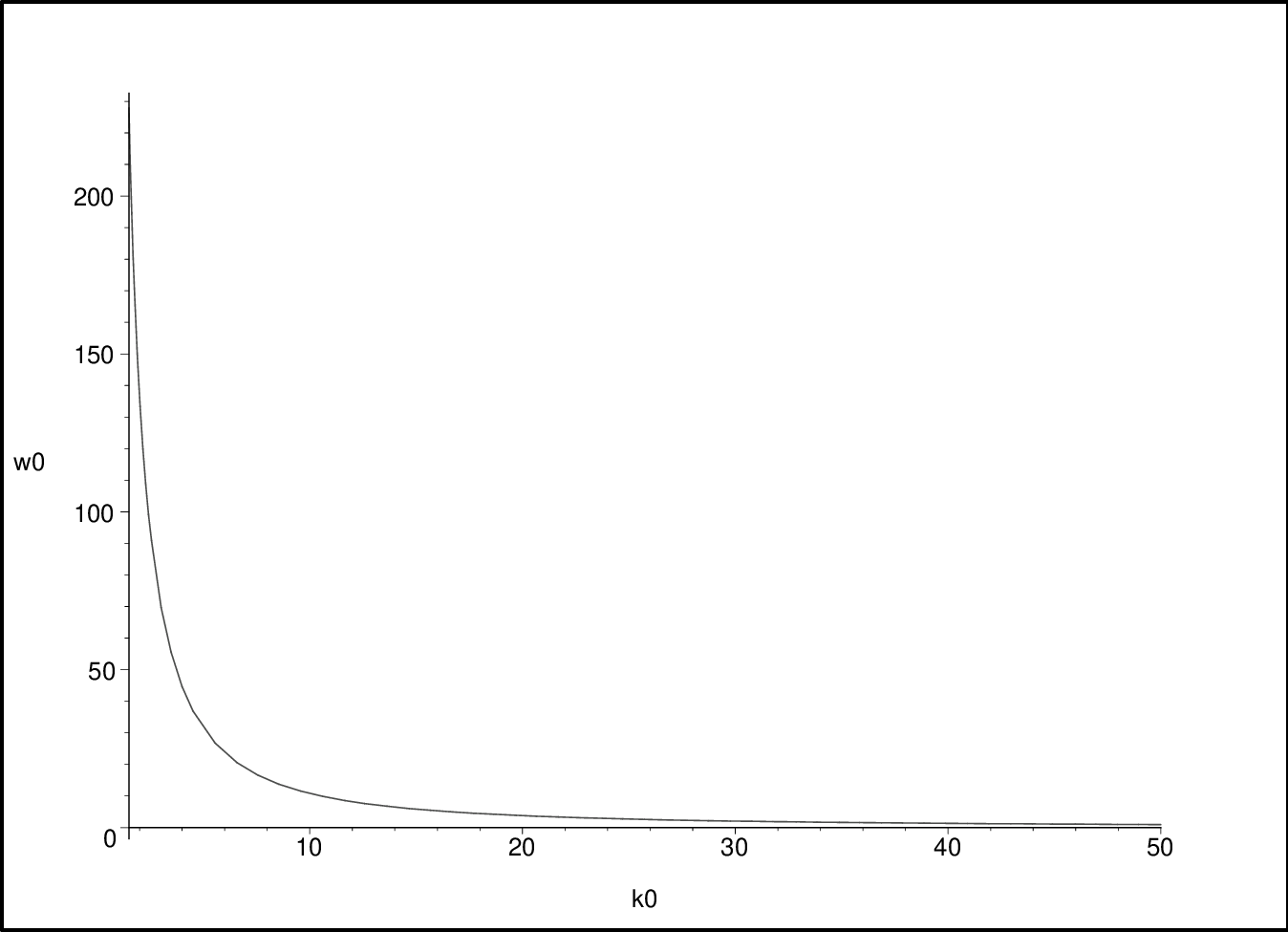,width=6cm,angle=270}}
\caption{$w_0$(kpc) as a function of $k_0$.      }
\label{figw0}
\end{figure}

\sub{Orders of magnitude and Solar System:} The solutions we have considered contain a length scale, $w_0$. This parameter was replaced in  (\ref{w0}) by the final speed $v_\infty$, which is a better observable.  It is however interesting to estimate the values of $w_0$ for a realistic situation. We set  $l \sim 10^6 kpc$ (cosmological length), and ${v_{\infty} \over c} \sim {1 \over 3} 10^{-3}$, for a typical situation with $v_\infty \sim 100 km/sec$. Fig. \ref{figw0} shows $w_0$ as a function of $k_0$.

For $k_{0} > 3 $, $w_0$ is equal to a few kpc.  This is a natural galactic scale. With an optimistic viewpoint one can thus assign to $w_0$ some physical meaning determined by the length of the object observed. In other words, the tensor $q_{\mu\nu}$ is a field whose natural length scale of variation is determined by the object.

Now, the natural dimensionless parameter which controls the corrections from flat space is ${w_0 \over l}$. If we believe that the value of $w_0$ is comparable to the object of study, then for Solar System experiments ${w_0 \over l}$ is too small, and the effects of $C^{\mu}_{\ \nu\rho}$ should not contribute.

~

\sub{Central density:} The central density associated to $C^{\mu}_{\ \nu\rho }$ diverges linearly, as the NFW profile (\ref{NFWp}). This can be seen by solving (\ref{g11}-\ref{g33}), for small values of $r$, as a series expansion. The series,
\begin{eqnarray}
  k(r) &=& k_0 - {\beta (k_0-1) \over w_0k_0 }\ r + {\cal O}(r^2) \\
  m^{{\scriptscriptstyle (1)}}(r) &=& -{w_0^2k_0^2 c^2 \over 2} \ r  + {\cal O}(r^2) \\
  u^{\scriptscriptstyle (1)}(r) &=& {w_0^2k_0^2c^2 \over 2} + {\cal O}(r^2)
\end{eqnarray}
solve (\ref{g11}-\ref{g33}) with the boundary condition $v(r)\rightarrow 0$ as $r\rightarrow 0$. With this solution at hand we can compute the behavior of the associated mass density,
\begin{eqnarray}
  4\pi G\, \rho(r) &=& {1 \over r^2}( r^2 \Phi' )'  \\
   &\simeq & { 2(k_0-1)w_0 c^2 \beta \over l^2\, r} + {\cal O}(1)
\end{eqnarray}
with a linear divergency, as anticipated.

\section{Eddington action, the equivalence principle and $g_{\mu\nu}=0$}

\label{Sec/g=0}

Our proposal for dark matter and dark energy is summarized in the action (\ref{I}). Once the action is written one can ``roll down" exploring its predictions and consequences by usual methods. This is what we have done so far. However, it is also interesting to ``climb up" and attempt a derivation, or at least a good motivation to include the Born-Infeld term in the gravitational action.

We start this section recalling a well-known effect. Consider a system of $N$ spins. If no external field is applied (and the temperature is not too small) the macroscopic average is $\langle \vec{S} \rangle = 0$.  On the contrary, in the presence of an external field, $H_{ext}$, the symmetry is broken, the spins align and produce a non-zero macroscopic average $\langle \vec{S} \rangle_{\vec{H}_{ext}}\neq 0$. It then follows that the total magnetic field felt by a charge $q$ is
\begin{equation}
\vec{H}_T = \vec{H}_{ext} + \langle \vec{S} \rangle_{\vec{H}_{ext}}.
\end{equation}
The orbit of the charge will obey the Lorentz equation with $\vec{H}_T$ not $\vec{H}_{ext}$. If we did not know about spins the contribution $\langle \vec{S} \rangle_{\vec{H}_{ext}}$ would be interpreted as a sort of `dark' magnetic field. If the temperature is below the Curie temperature, the external field could be removed and the spins remain in their `ordered' state with $\langle \vec{S} \rangle_0 \neq 0$.

Let us now describe an analog of this effect in the theory of gravity.  Topological manifolds are invariant under the full diffeomorphism group. Riemannian manifolds are invariant only under the subgroup of isometries of the metric. The state $g_{\mu\nu}=0$ represents the unbroken state of general relativity \cite{Witten88}, and the  introduction of a metric breaks the symmetry. The natural geometrical analog of the external field $\vec{H}_{ext}$ is the metric tensor $g_{\mu\nu}$. (See \cite{Horowitz,Giddings,Guendelman} for other discussions on the state $g_{\mu\nu}=0$, and \cite{Witten07} for a recent critical viewpoint.)

We shall treat the metric as an external field which can be switched on and off\footnote{In this picture, the big-bang could be understood as a smooth transition from a manifold without metric into a Riemanian manifold.}. Our first goal is to explore fields that can be defined in the absence of a metric. The simplest example is given by a connection $C^{\mu}_{\ \nu\rho}(x)$.  In fact, Eddington introduced a purely affine theory a long time ago \cite{Eddington},
\begin{equation}\label{edd0}
I_0[C] = \kappa \int d^4 x\, \sqrt{K_{\mu\nu}(C)}
\end{equation}
where $K_{\mu\nu}$ is the curvature associated to the connection $C^{\mu}_{\ \nu\rho}(x)$ (see Eqn. (\ref{Kmn})).  This action is invariant under spacetime diffeomorphism and yields second order differential equations for the field $C^{\mu}_{\ \nu\rho}$.  Eddington action was extensively studied as a purely affine theory of gravity, and also as a possible unification of gravity and electromagnetism \cite{Eddington,Poplawski}. We take here a different interpretation and let the field $C^{\mu}_{\ \nu\rho}$ be an independent degree of freedom.

We now turn on the external field $g_{\mu\nu}$ and study the effects of both $g_{\mu\nu}$ and $C^{\mu}_{\ \nu\rho}$ on particles. The first problem is to determine  the action for the coupled system.  We do not want to introduce ghost or higher derivatives. The action  (\ref{edd0}) is already free of anomalies. So we start by adding the standard Einstein-Hilbert action for $g_{\mu\nu}$ and consider
\begin{equation}
\int d^4 x  \left( \sqrt{g}R  + \kappa \sqrt{K_{\mu\nu}}\  \right).
\end{equation}
With this action, the fundamental fields $g_{\mu\nu}$ and $C^{\mu}_{\ \nu\rho}$ are decoupled. To make the theory more interesting we add interactions. The most attractive  theory (although not unique) having second order field equations is the Einstein-Born-Infeld action introduced in Eq. (\ref{I}).

An important point now is to define the geodesic equation for the coupled system. In the presence of a metric $g_{\mu\nu}$ there is a natural affine connection $\Gamma^{\mu}_{\ \nu\rho}$ represented by the Christoffel symbol,
\begin{equation}\label{chr}
\Gamma^\mu_{\ \nu\rho} = {1 \over 2}g^{\mu\sigma} ( g_{\sigma\nu,\rho} + g_{\sigma\rho,\nu} - g_{\nu\rho,\sigma}).
\end{equation}
The question is, should geodesics be defined with respect to $C^{\mu}_{\ \nu\rho}$, $\Gamma^{\mu}_{\ \nu\rho}$, both?  In order to comply with the equivalence principle we shall postulate that particles only couple to the metric and not to the connection $C^{\mu}_{\ \nu\rho }$.  The geodesic equation then take the usual form
\begin{equation}\label{geo}
\ddot x^{\mu} + \Gamma^{\mu}_{\ \alpha\beta} \dot x^\alpha \dot x^\beta=0,
\end{equation}
where $\Gamma^{\mu}_{\ \alpha\beta}$ is the Christoffel symbol (\ref{chr}).  Observe that the metric satisfies the equations (\ref{ee}) and is coupled to the field $C^{\mu}_{\ \nu\rho }$. In this sense, $C^{\mu}_{\ \nu\rho }$ does contribute to $g_{\mu\nu}$ and indirectly affects the motion of particles. This is how the field $C$ can explain flat rotation curves.

Now, the analogy with spin systems can be pushed a little bit further.  We have seen in the cosmological analysis that for large times the system approaches the de-Sitter solution (see Sec. \ref{SecdeSitter}), and in particular the metric $q_{\mu\nu}$ becomes proportional to $g_{\mu\nu}$, $q_{\mu\nu}\rightarrow \lambda g_{\mu\nu}$. One can interpret this fact as analogous to the alignment of spins along the direction of the applied field, $\langle \vec{S} \rangle  \rightarrow \lambda \vec{H}_{ext}$. Of course, to support this interpretation one would need to consider generic initial conditions. This will be analyzed elsewhere.

Finally, recall that when the external  magnetic field is removed, spins can have a spontaneous non-zero average $\langle \vec{S} \rangle$, and this vector generates forces on charged particles.  Is there a gravitational analogue to this effect? The gravitational force is measured by the connection (\ref{chr}), entering in the geodesic equation. The external field is the metric. Now, as the metric is removed, the Christoffel connection becomes ${0 \over 0}$, with the same scaling weight in the numerator and denominator. For a large class of paths the limit is a finite function.  Since the only connection available at $g_{\mu\nu}=0$ is $C^{\mu}_{\ \nu\rho}$, it is tempting to conjecture that $\Gamma^{\mu}_{\ \nu\rho} \rightarrow C^{\mu}_{\ \nu\rho}$, as the metric is removed. In this way, the geodesic equation has a non-trivial limit when the metric vanishes, and particles will feel `forces'.  These are not forces in the usual sense because there is no metric.  (Although note that a geodesic equation, defined by parallel transport, can be introduced without a metric.)  The limit $g_{\mu\nu}\rightarrow 0$ was the key ingredient employed in \cite{B} for a different approach to understand dark matter as an effect associated to a topological manifold. To make these ideas precise a theory describing the process $g_{\mu\nu}\rightarrow 0$ is necessary. We hope to come back to this interpretation in the future.

\section{Conclusions}

Dark matter and dark energy have unique properties and their understanding in one of the most crucial problems faced by theoretical physics today.   Dark matter does not interact with normal matter and this property has motivated us to look for fields which have this property somehow ``built in". We have explored gravity coupled to connection $C^{\mu}_{\ \nu\alpha}$ field with a Born-Infeld action.

This theory comply with the main background properties normally attributed to dark matter and dark energy. First, the evolution of the scale factor in cosmological models has the right time dependence interpolating between pressureless matter and a cosmological constant.

At galactic scales dark energy is less relevant but dark matter still plays an important role. By an approximation valid for distances much smaller to the Hubble radius we have solved the equations of motion for spherical objects and find the expected rotation curves.  These curves satisfy the basic asymptotic flatness observed in galaxies providing new support for this proposal.

We have left several topics for the future. The stability of this theory and the study of primordial fluctuations are important to determine the CMB anisotropies. This will be reported in \cite{BFS}.  On galactic scales a systematic fit with observational curves is necessary. This issue is presently under study and will be reported in \cite{BRR}.

\section{Appendix. Derivation of the equations of motion}

The fields which are varied in the action (\ref{I}) are the metric $g_{\mu\nu}$ and the connection $\Gamma^{\mu}_{\ \nu\rho}$. The equations of motion for the metric follow by a straightforward variation of the action. The result is
\begin{equation}\label{1}
G_{\mu\nu} = \sqrt{{|g_{\mu\nu}- l^2 K_{(\mu\nu)}| \over |g_{\mu\nu}| }} \ g_{\mu\alpha}\left({1 \over g-l^2 K }\right)^{\alpha\beta} g_{\beta\nu} + 8\pi G \, T^{{\scriptscriptstyle (m)}}_{\mu\nu}
\end{equation}
This equation can be drastically simplified by using the equation of motion for the connection $\Gamma^{\mu}_{\ \nu\rho}$. This equation is derived in two steps.  First, since the action only depends on the curvature $K_{\mu\nu}$ once can compute the variation using the chain rule,
\begin{equation}
{\delta I \over \delta \Gamma^{\mu}_{ \nu\rho}} = \int {\delta I \over \delta K_{(\alpha\beta)}} \, {\delta K_{(\alpha\beta)} \over \delta \Gamma^{\mu}_{\nu\rho}}
\end{equation}
Just like in Eddington \cite{Eddington} theory one finds  by direct variation that the combination
\begin{equation}\label{K00}
\sqrt{q}{q}^{\mu\nu}  \equiv -{1 \over \alpha}\sqrt{|g_{\mu\nu} - l^2 K_{\mu\nu}|} \left({1 \over g-l^2 K }\right)^{\mu\nu}
\end{equation}
satisfies
\begin{equation}
D_\rho(\sqrt{q}{q}^{\mu\nu})=0
\end{equation}
where $D_\rho$ is the covariant derivative built with the connection $\Gamma_{{\scriptscriptstyle 0}}$.  Since $\Gamma^{\mu}_{\ \nu\rho}$ is symmetric, this equation imply
\begin{equation}
\Gamma^{\mu}_{ \nu\rho} = {1 \over 2} q^{\mu\alpha} ( q_{\alpha\nu,\rho} + q_{\alpha\rho,\nu} - q_{\nu,\rho,\alpha} )
\end{equation}
We thus write $\Gamma^{\mu}_{\ \nu\rho}$ in terms of $q_{\mu\nu}$. The equation (\ref{K00})  now depends only on $q_{\mu\nu}$. Taking the determinant at both sides, and inverting one readily derives (\ref{Ke}).

The final simplification follows by noticing that the right hand side of (\ref{1}) contains $\sqrt{q}q^{\mu\nu}$.  Thus, using (\ref{K00}), Eq. (\ref{1}) is transformed into (\ref{ee}).

The analysis of these equations is greatly simplified by using the bi-metric formalism \cite{andy}.

~

~

\section{Acknowledgements}

The author would like to thank S. Carlip, P. Ferreira, A.Gomberoff, M. Henneaux, A. Reisenegger, D. Rodrigues,  N. Rojas, C. Skordis and S. Theisen for useful comments and discussions.  The author was partially supported by Fondecyt Grants (Chile) \#1060648 and \#1051084.

 \end{document}